# Design of multifunctional metamaterials using optimization

Ewan Fong[1], Sadik L Omairey[1,2], Peter D Dunning[1*]

[1]*School of Engineering, University of Aberdeen, Aberdeen, UK*
[2]*Brunel Composites Centre, College of Engineering, Design and Physical Sciences, Brunel University London, London, UK*

**Abstract**

This paper explores the use of optimization to design multifunctional metamaterials, and proposes a methodology for constructing a design envelope of potential properties. A thermal-mechanical metamaterial, proposed by Ai and Gao (2017), is used as the subject of the study. The properties of the metamaterial are computed using finite element-based periodic homogenization, which is implemented in Abaqus utilizing an open-source plugin (EasyPBC). Several optimization problems are solved using a particle swarm-based optimization method from the pyOpt package. A series of constrained optimization problems are used to construct a design envelop of potential properties. The design envelope more fully captures the potential of the metamaterial, compared with the current practice of using parametric studies. This is because the optimizer can change all parameters simultaneously to find the optimal design. This demonstrates the potential of using an optimization-based approach for designing and exploring multifunctional metamaterial properties. This proposed approach is general and can be applied to any metamaterial design, assuming an accurate numerical model exists to evaluate its properties.

Keywords: metamaterial; auxetic material; non-positive thermal expansion coefficient; periodic homogenization; particle swarm optimization

[*]Corresponding author

Address: Fraser Nobel Building, King's College, Aberdeen, AB24 3UE, UK

Email: peter.dunning@abdn.ac.uk

# 1. Introduction

Metamaterials are engineered materials that have properties not usually found in nature, due mainly to their structure, rather than their material composition. Early research in metamaterials focussed on electromagnetic properties, such as materials with negative permittivity and permeability (Engheta and Ziolkowski 2006), but the idea has been extended to other areas, including mechanical (Huang and Chen 2016; Ren et al. 2018), thermal (Sklan and Li 2018) and thermal-mechanical properties. Auxetic materials are mechanical metamaterials with a negative Poisson's ratio, such that the material expands in the transverse direction when a tensile strain is applied in the longitudinal direction and vice versa. Many auxetic materials have been developed since the seminal work by Lakes (1987) on auxetic foams. These include lattice structures, rotating polygons, chiral structures, crumpled and perforated sheets (Ren et al. 2018). Auxetic materials have been shown to have several useful engineering properties, such as improved indentation resistance, energy absorption and fracture resistance (Huang and Chen 2016; Ren et al. 2018). Thus, they have found many applications, such as medical devices (e.g. stents), protective devices and smart sensors and filters (Ren et al. 2018).

Metamaterials with a non-positive coefficient of thermal expansion (CTE) either contract with temperature increase (negative CTE) or remain the same size (zero CTE). It has been shown that cellular metamaterials composed of two materials with different CTE values and some void space can obtain zero and negative CTE values (Lakes 1996; Sigmund and Torquato 1996). Non-positive CTE metamaterials have potential use in temperature-sensitive applications, such as sensors, thermo-mechanical actuators and structures subject to thermal shock (Huang and Chen 2016).

Many studies focus on designing metamaterials for a single novel property. However, some studies aim to design multifunctional metamaterials that have two or more novel properties. For example, there have been several recent studies designing metamaterials with both negative Poisson's ratio and non-positive CTE. Grima et al. (2007) proposed a 2D lattice metamaterial composed of connected triangles, where one side of a triangle is made from a material with different CTE than the other two sides. Ha et al. (2015) showed experimentally that a bimetallic auxetic chiral metamaterial can also have negative CTE and that Poisson's ratio and CTE are independent. Ai and Gao (2017) proposed four bi-material lattice metamaterials, based on 2D star-shaped re-entrant structures. These were analysed using finite element analysis (FEA) based periodic homogenization and one of the proposed designs obtained both auxetic behaviour and non-positive CTE. This work has been extended to 3D metamaterial lattices (Ai and Gao 2018), demonstrating the first true 3D lattice metamaterial that can be both auxetic while also having non-positive CTE. Raminhos et al. (2019) experimentally investigated one of the metamaterial designs of Ai and Gao (2017) (with polymeric materials, instead of metallic) and confirmed the auxetic and negative CTE properties.

Other bi-material re-entrant 2D lattice metamaterials that have both auxetic behaviour and non-positive CTE have been proposed and analysed numerically (using FEA). Ng et al. (2017) used re-entrant triangles, Wei et al. (2018) used a combination of single material re-entrant triangles and the bi-material triangles, first proposed by Grima et al. (2007), and Li et al. (2019) proposed star-square structures, which have some similarities with the designs of Ai and Gao (2017), but with an additional outer box (square) for each cell. The auxetic behaviour of the star-square designs was also confirmed experimentally.

It has been speculated that multifunction metamaterials (such as those with auxetic behaviour and non-positive CTE) could have a wide range of medical, aerospace and defence applications, as they can help promote the development of multi-functionality and multi-purpose devices (Huang and Chen 2016; Ren et al. 2018). Ng et al. (2017) suggest these materials may be used for electronic sensors and actuators and the development of composite materials with extreme mechanical and thermal load-bearing capacity. However, this area of research has only recently seen activity and it has been noted that further investigation is required on the design and optimization of multifunctional metamaterials (Ren et al. 2018; Raminhos et al. 2019).

The studies mentioned above that propose and design metamaterials for both auxetic behaviour and non-positive CTE only use parametric studies to explore the potential range of properties, where only one variable is changed at a time, whilst others remain constant. Thus, this parametric approach does not fully explore the design space and the full range of potential properties. Optimization methods, particularly stochastic methods, can better explore the design space by simultaneously changing all design variables.

Thus, the aim of this study is to explore the use optimization methods to design a multifunctional metamaterial. In particular, a methodology is proposed for constructing a design envelope for a multifunctional metamaterial by solving a series of constrained optimization problems. The chosen case study is metamaterial that can exhibit both negative Poisson's ratio and a range of CTE values, including zero and negative values. One of the 2D bi-metallic lattice star-shaped structures proposed by Ai and Gao (2017) is used in the case study, which was previously analysed using parametric studies only. Section 2 introduces the methodology used, including FEA-based periodic homogenization and optimization, Section 3 presents results, which is followed by discussion and conclusions in Sections 4 and 5.

## 2. Methodology

*2.1 Metamaterial description*

As stated previously, the metamaterial considered in this study is based on a design proposed by Ai and Gao (2017). Their study considered a set of four different bi-material lattice structures with the goal of creating a metamaterial that can exhibit both non-positive CTE ($\alpha \leq 0$) and negative Poisson's Ratio ($v < 0$). In particular, it was found that one of the lattice structures (Fig. 1) provided promising results; where both negative Poisson's Ratio and non-positive CTE values can be exhibited. Additionally, the study considered three constituent materials; Aluminium Alloy, Steel and Invar, resulting in three different material pairings. However, it was found that the Aluminium-Invar pair led to a metamaterial exhibiting non-positive CTE over a wide range of design parameters. Thus, the Aluminium-Invar material pairing is chosen in this study to provide the optimizer with a larger design space for both non-positive CTE along with negative Poisson's Ratio. Properties of the constituent materials are summarized in Table 1.

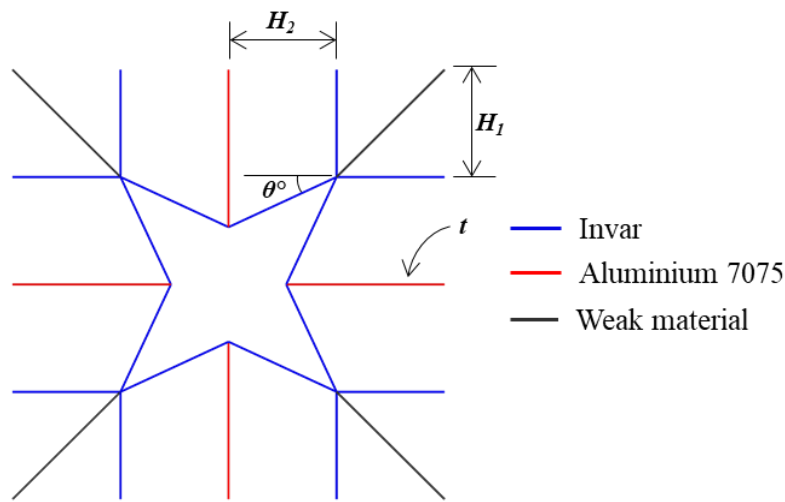

Figure 1. Star-shaped lattice metamaterial concept.

Table 1. Temperature dependent material properties of the constituent materials.

| Material | Young's Modulus, E (GPa) | CTE, $\alpha$ (K$^{-1}$) | Poisson's Ratio, $v$ | Temperature (°C) |
|---|---|---|---|---|
| Al 7075 | 71 | 23.0 x 10$^{-6}$ | 0.33 | 20 |
|  | 66 | 24.3 x 10$^{-6}$ | 0.33 | 200 |
| Invar | 144 | 1.1 x 10$^{-6}$ | 0.29 | 20 |
|  | 135 | 2.5 x 10$^{-6}$ | 0.29 | 200 |

The studied lattice metamaterial is shown in Fig. 1. As stated previously, this structure incorporates the ideas of a bi-material lattice structure in its design to create a material that can exhibit negative Poisson's Ratio and near-zero CTE. When the material is heated the difference in CTE values of the constituent materials causes the overall structure to expand inwards into its own void space, rather than expanding externally. Additionally, when this material is strained mechanically, the structure expands in both the axial and lateral directions. This is due in part to the shape of the lattice structure, but also the difference in Young's Modulus between the two constituent materials. Note that the lattice structure is square symmetric: $v_{12} = v_{21} = v$, and $α_1 = α_2 = α$.

However, it must be noted that the lattice structure used in this study has one difference to the one used by Ai and Gao (2017); as extra beams in the corners are incorporated to make it compatible with EasyPBC for periodic boundary condition application (see Section 2.2. for more discussion). To ensure these beams have a negligible impact on the overall properties, a fictitious "weak material" is used. The weak material has a low Young's modulus of E = 1 kPa and a CTE similar to that of both Invar and Aluminium of $α = 10^{-6}$ K$^{-1}$. Therefore, despite this small modification, the structure behaves almost identically to that of Ai and Gao (2017), validated by comparing the material properties calculated in this study to those reported by Ai and Gao.

The FE package Abaqus is used to model and analyse the lattice metamaterial structure. Beam elements are used as the members in the lattice are long and thin (beam elements were also used by Ai and Gao (2017)). Two-node Timoshenko beams (element type B31) are used in this study. Automatic meshing is used, with a seed size of 0.085 times the RVE edge length, which creates meshes with between 150-200 elements. As mentioned above, the results using this model were compared against those reported in Ai and Gao (2017) and found to be almost identical.

*2.2 Periodic homogenization*

FEA-based periodic homogenization is a widely-used method to estimate the properties of multi-phase materials by analysing a representative volume element (RVE) to reduce unnecessary complexity and replace the heterogeneous structure of the material (Geers et al. 2010). To accurately simulate an RVE, its opposite boundaries must have equal displacements to form a periodic continuous material. This can be done by prescribing specific displacement constraint conditions at the RVE's boundary nodes. For this purpose, an open-source Abaqus plugin, EasyPBC (Omairey et al. 2019), is used to automatically generate the periodic boundary conditions, Eqs. 1.

$$U_{Right}^{x} - U_{Left}^{x} = \Delta^{x} \tag{1a}$$

$$U_{Right}^{y} - U_{Left}^{y} = 0 \tag{1b}$$

$$U_{Top}^{x} - U_{Bottom}^{x} = 0 \tag{1c}$$

$$U_{Top}^{y} - U_{Bottom}^{y} = \Delta^{y} \tag{1d}$$

where $U^x$ and $U^y$ are the displacement components along the X and Y directions, shown in Fig. 2. $Δ^x$ and $Δ^y$ are the extension of the RVE in the X and Y directions, respectively, and used to apply or measure strain values (see Sections 2.2.1 and 2.2.2). The version of EasyPBC used in this study requires the RVE to have corners nodes to generate the required periodic boundary conditions, which are absent in the selected lattice structure. This is addressed by adding extra beams made from a fictitious weak material that extends from the Invar to form the required corner nodes, as shown in Fig. 2 and discussed above in Section 2.1.

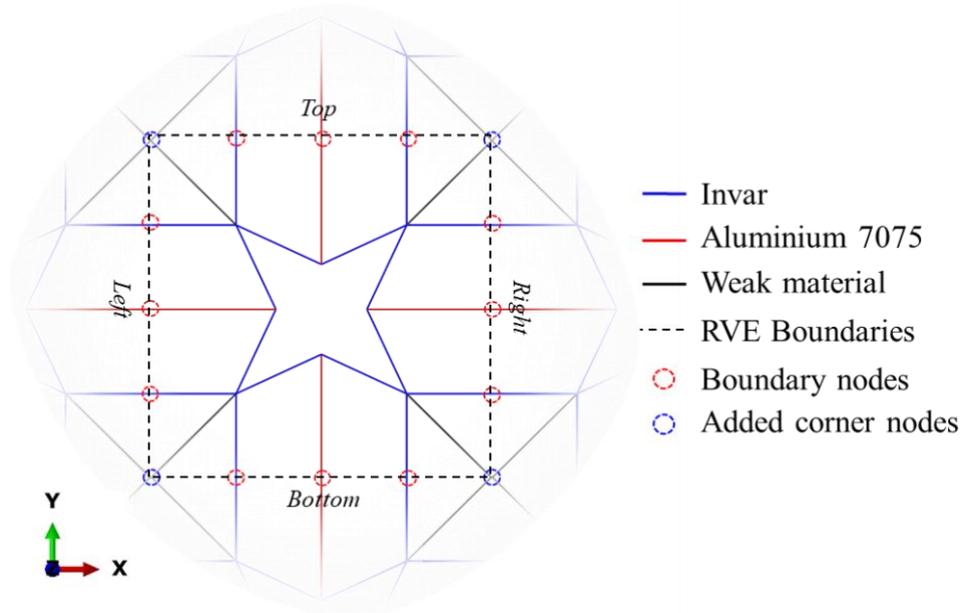

Figure 2. RVE and periodic boundary conditions of the studied metamaterial.

Note that periodic homogenization is based on the separation of scales, such that the actual dimensions within the micro-scale lattice structure are only relative to the size of the RVE (which can be chosen arbitrarily) and cannot be related to real physical dimensions. Hence, length units for the geometric dimensions of the lattice structure are not used in this paper, as they have no meaning.

*2.2.1 Computing Poisson's ratio*
To calculate Poisson's ratio, EasyPBC applies strains to the RVE by prescribing a displacement in the X direction only (as a square symmetric structure is investigated). Hence, $Δ^x$ in Eq. 1a is the axial extension and is prescribed, whereas $Δ^y$ in Eq. 1d is the transverse extension and a free variable computed by the FEA. Using these values, Poisson's ratio is calculated by Eq. 2.

$$v = -\frac{transverse\ strain}{axial\ strain} = -\frac{\frac{Δ^y}{L}}{\frac{Δ^x}{L}} = -\frac{Δ^y}{Δ^x} \tag{2}$$

*2.2.2 Computing linear thermal expansion coefficient*

The linear thermal expansion coefficient is calculated using Eq. 3.

$$\alpha = \frac{\Delta^L}{L \Delta^T} \qquad (3)$$

where α is the linear thermal expansion coefficient, L the original length, ΔL the change in length of the RVE, and $\Delta^T$ the change in temperature. In this case, both $\Delta^x$ and $\Delta^y$ in Eqs. 1 are free variables computed by the FEA and both are equal to ΔL (due to the square symmetry of the lattice structure). A uniform temperature increase from 20°C to 200°C ($\Delta^T$ = 180°C) is applied to the RVE and $\Delta^L$ calculated from nodal displacement data, which is then used to compute the thermal expansion coefficient using Eq. 3.

*2.3 Optimization*

This study considers three main optimization objectives: minimization of Poisson's ratio (to achieve auxetic behaviour), minimization of CTE squared (to achieve near-zero CTE) and minimization of CTE (to obtain negative CTE). However, as CTE has very small values, which may affect optimizer convergence, it is replaced by the Normalized CTE (NTCE), which is defined here as the CTE of the metamaterial divided by the CTE of Aluminium ($\alpha_{Al}$ = 23×10$^{-6}$). This definition was also used by Ai and Gao (2017). A key part of the proposed methodology is to create a design envelope of possible properties for the studied metamaterial. This is achieved by solving a series of constrained optimization problems, namely minimizing NCTE, subject to different constraints on Poisson's ratio, and minimizing Poisson's ratio with different constraints on NCTE. The design variables used in the optimization are the four in-plane geometric dimensions: $H_1$, $H_2$, θ and *t* (as shown in Fig. 1). Note that Poisson's ratio and CTE are independent of the out-of-plane thickness of the structure. Thus, the out-of-plane thickness is not subject to optimization in this study. The upper and lower limits of the design variables are specified in Table 2 (remembering that lengths in the micro-structure only have relative meaning, hence no units are given – see discussion in Section 2.2).

Table 2. Design variable lower and upper limits.

| Variable | Lower limit | Upper limit |
|---|---|---|
| $H_1$ | 5.0 | 100.0 |
| $H_2$ | 5.0 | 100.0 |
| θ (°) | 5.0 | 40.0 |
| *t* | 0.5 | 5.0 |

The framework used to solve the optimization problems is implemented in Python, as this can easily interface with the Abaqus environment. A Python script is created that automatically generates the lattice structure geometry for a given set of design variables and applies all necessary analysis details, such as material properties and beam section properties. EasyPCB is then called to apply periodic boundary conditions and define the applied displacement, or temperature increase.

The script is interfaced with the pyOpt optimization package (Perez et al. 2012) to solve the optimization problems defined above. The package includes several optimizers, including both gradient-based and stochastic gradient-free methods. In this study, stochastic optimizers are used as they are suitable for exploring the design space, without getting trapped in local minima. Two optimizers are considered: Augmented Lagrangian Particle Swarm Optimizer (ALPSO) (Jansen & Perez, 2011) and Augmented Lagrangian Harmony Search Optimizer (ALHSO) (Geem et al. 2001). Both methods use an augmented Lagrangian approach to handle constrained problems. Full details of these optimizers are omitted for brevity but can be found in references above. The performance of these two optimizers for the problems studied here is compared in Section 3.1. The default settings for both optimizers are used throughout this study.

## 3. Results

The framework outlined above for analysing and optimizing the multifunctional metamaterial is now used to explore the extreme properties that can be achieved and the range of possible properties by constructing a design envelope.

### 3.1 Comparison of optimizers

First, the two considered optimizers are compared over three different optimization objectives, minimizing Poisson's Ratio, minimizing CTE and near-zero CTE, in order to find the most suitable optimizer for this study. The results are summarized in Table 3, including the optimal design, number of design evaluations (i.e. number of separate FEA computations) and objective function values. In all cases the default options from the pyOpt package are used.

Table 3. Comparison of optimizer performance.

| Objective | Minimize Poisson's ratio | | Minimize NCTE | | Near-zero NCTE | |
|---|---|---|---|---|---|---|
| Optimiser | ALHSO | ALPSO | ALHSO | ALPSO | ALHSO | ALPSO |
| $H_1$ | 100 | 100 | 97.64 | 100 | 42.34 | 81.82 |
| $H_2$ | 13.34 | 13.34 | 61.74 | 25.01 | 89.46 | 43.73 |
| θ (°) | 23.85 | 23.85 | 39.38 | 40.00 | 14.20 | 9.86 |
| $t$ | 0.5 | 0.5 | 0.69 | 0.50 | 4.95 | 1.10 |
| Evaluations | 1930 | 1240 | 1581 | 1240 | 273 | 280 |
| $v$ | -0.386 | -0.386 | -0.228 | -0.291 | -0.132 | -0.201 |
| NCTE | -0.246 | -0.246 | -0.607 | -0.647 | -0.0005 | -0.0002 |
| CTE ($K^{-1}$) | -5.66x$10^{-7}$ | -5.66x$10^{-7}$ | -1.40x$10^{-6}$ | -1.49x$10^{-6}$ | -1.15x$10^{-8}$ | -4.60x$10^{-9}$ |

For minimizing Poisson's ratio, both optimizers found the same design, although ALPSO took 1240 evaluations to find this solution, whereas ALHSO took 1930 (a 50% increase). Additionally, when comparing both optimizers on minimizing the value of CTE, ALPSO found a better solution (lower CTE) using fewer evaluations than ALHSO. Finally, both performed similarly when optimizing for near-zero CTE, as both found solutions with CTE values less than one-thousandth of the constituent materials, which can be considered near-zero. However, the solutions are different, which indicates that there are multiple designs of the lattice structure that can achieve near-zero CTE. In summary, ALPSO generally performed better than ALHSO (using the default options) for the problems studied here, by either obtaining the solution using fewer evaluations or finding a better optimum. Thus, ALPSO is used as the optimizer for the remainder of this article. It should be noted that both optimizers may perform better if optimization parameters are tuned to the problem, but this is beyond the scope of this article.

*3.2 Minimization of Poisson's ratio*

In Section 3.1, a solution is found for minimizing Poisson's Ratio, with: $H_1$ = 100, $H_2$ = 13.34, θ = 23.85° and $t$ = 0.5, giving a Poisson's Ratio of $v$ = -0.386. Fig. 3 shows both the undeformed (unstrained) and deformed (strained) structure. It must be noted that Fig. 3 is a magnified view focussed on the centre of the lattice structure, in order to highlight the mechanism that results in a negative Poisson's Ratio.

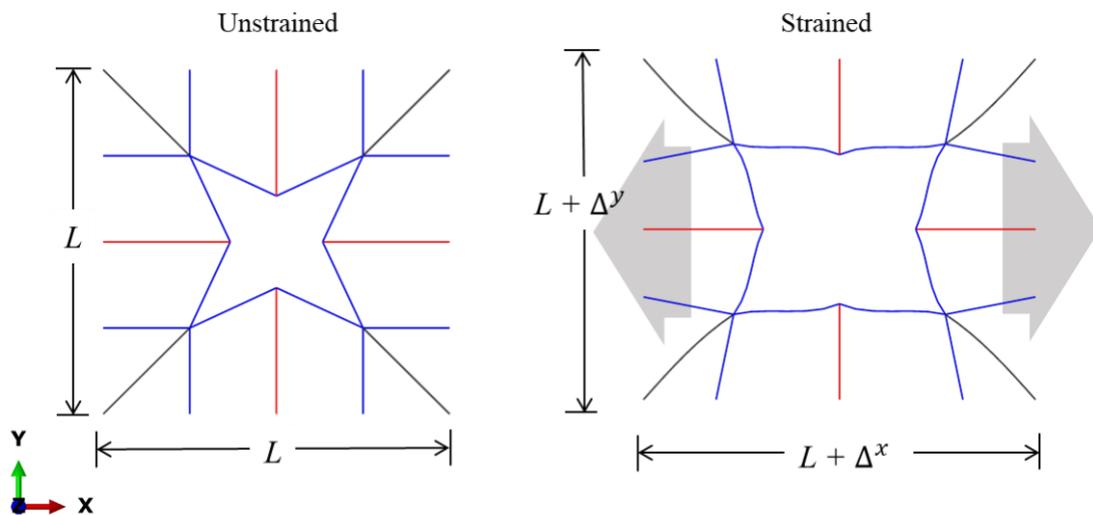

Figure 3. Magnified view of metamaterial structure for minimum Poisson's ratio.

The deformed state of the structure shows that when a strain is applied in the positive x-direction, there is a decrease in θ, resulting in the Aluminium (red) struts moving away from the centre. As discussed above, this behaviour is due both to the shape of the lattice structure and the difference in Young's Modulus values of the two constituent materials. This optimization result shows that a large difference in $H_1$ and $H_2$, with thickness, $t$, at the lower bound, exaggerates this mechanism to produce a metamaterial with a reasonably large negative Poisson's ratio.

*3.3 Near-zero thermal expansion coefficient*

The optimal solution for near-zero CTE found by ALPSO (Table 3) has design variables of: $H_1$ = 81.82, $H_2$ = 43.73, θ = 9.86° and $t$ = 1.1, giving a NCTE of αn = -2×$10^{-4}$, or a CTE of α = -4.6×$10^{-9}$ $K^{-1}$. The undeformed and deformed structures are shown in Fig. 4.

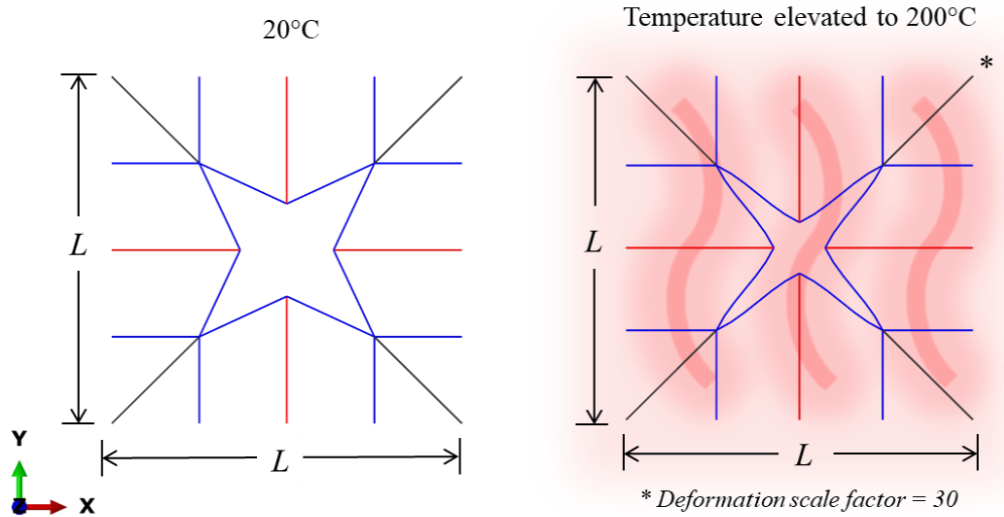

Figure 4. Undeformed and deformed view of a metamaterial with near-zero CTE.

Similar to the solution for minimum Poisson's ratio (Fig. 3), it can be seen in Fig. 4 that there is a distinct change in θ. However, under a uniform temperature increase, the change in θ is positive. This is due to the difference in CTE of the two constituent materials, as Aluminium has a larger CTE ($α_{Al}$ = 23 x $10^{-6}$ $K^{-1}$) compared with Invar ($α_{Invar}$ = 1.1x $10^{-6}$ $K^{-1}$). Thus, when this material is subjected to a temperature increase, the Aluminium (red) struts increase in length more than the Invar (blue) struts. This difference in CTE causes the inner structure to "collapse" inwards, as the red Aluminium struts push the blue Invar struts towards the centre. As a result, while these Invar struts do expand, they expand towards the void at the centre of the structure. This allows the metamaterial to maintain its volume, resulting in a near-zero CTE.

The design and mechanism for minimum Poisson's ratio shown in Fig. 3 is different from that for near-zero CTE shown in Fig. 4. This highlights the difficulty in obtaining a multifunctional metamaterial with both negative Poisson's ratio and near-zero CTE. However, Table 3 shows that ALHSO obtained a different solution for near-zero CTE, suggesting that there are potentially many designs that can achieve near-zero CTE. Hence, the range of potential material properties for the lattice structure studied in this paper is explored further in Section 3.5.

*3.4 Minimization of thermal expansion coefficient*

ALPSO is now used to minimize CTE. The solution has dimensions of: $H_1$ = 100, $H_2$ = 25.01, θ = 40° and $t$ = 0.5, giving a metamaterial with a NCTE of $α_n$ = -0.647 and a CTE of α = -14.9×10$^{-6}$ K$^{-1}$.

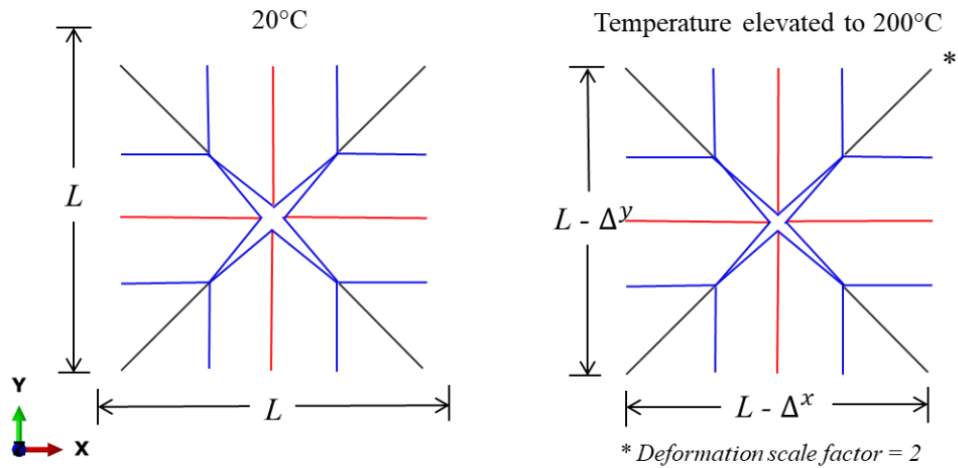

Figure 5. Magnified view of the solution for minimum CTE.

Similar to the solution for near-zero CTE shown in Fig. 5, the temperature increase creates a positive angle change in the structure. However, for this design, there is a much greater angle increase compared with those optimized for near-zero CTE. As a result, this structure contracts when subjected to a temperature increase.

*3.5 Design envelope*

Further optimizations are performed to explore the range of possible properties of the multifunctional metamaterial and construct a design envelope. The first set of optimizations aims to minimize Poisson's Ratio while constraining NCTE to be larger than a prescribed value. The second set also minimizes Poisson's Ratio, whilst constraining NCTE to be smaller than a prescribed value. Then, NCTE is minimized, whilst constraining Poisson's ratio to be greater than a prescribed value. Finally, the optimizer is used to maximize Poisson's ratio and then NCTE. The complete set of results are plotted in Fig. 6.

The results show a wide range of solutions for both negative Poisson's Ratio and non-positive NCTE values; covering ranges of: $-0.386 \leq \nu \leq 0.0$ and $-0.647 \leq α_n \leq 0.0$, or $-14.9\times10^{-6} \leq α \leq 0.0$ K$^{-1}$. The approximate range of these properties reported by Ai and Gao (2017) using parametric studies are smaller: $-0.19 \leq \nu \leq 0.0$ and $-0.6 \leq α_n \leq 0.0$, or $-13.8\times10^{-6} \leq α \leq 0.0$ K$^{-1}$. Thus, this example highlights how optimization can be used to explore the full design space of a multifunctional metamaterial concept, revealing the range of possible material properties. Then, using the design envelope as a guide, metamaterial properties can be tailored towards specific applications by using an optimizer to minimize, maximize and possibly constrain properties to achieve a desired behaviour.

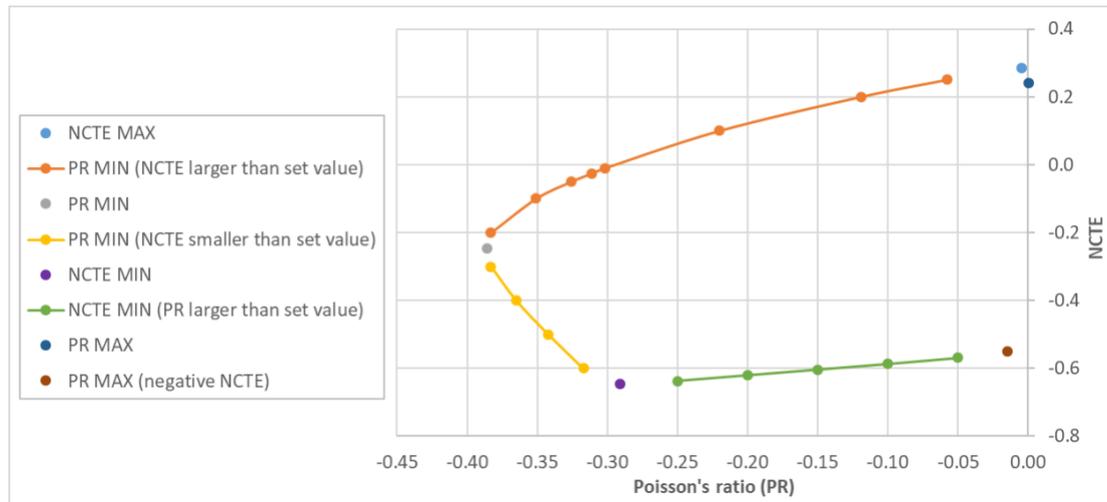

Figure 6. Design envelope of the multifunctional metamaterial.

Further insight is gained by examining the variation of optimal design parameters along the edge of the design envelope. The full set of results is summarized in Appendix A. For example, when minimizing Poisson's ratio, with constraints on NCTE, there are clear trends in design parameters. As NCTE is decreased, $H_2$ increases, whilst $H_1$ remains at (or near) its upper limit, θ also increases towards its upper limit and $t$ quickly decreases to its lower limit. Then, as NCTE is minimized, with constrains on Poisson's ratio, these trends continue, until $H_2$ reaches its upper limit, when $H_1$ starts to decrease.

It is also possible to construct parts of the design envelop using a multi-objective optimization approach. For example, the yellow part in Fig. 6 could be constructed by solving a multi-objective problem to minimize both Poisson's ratio and NCTE. However, multiple problems will still need to be solved to construct the full envelop and the constrained optimization approach used here proved effective and efficient for this example problem.

## 4. Discussion

The results in this study demonstrate how an optimization approach can be used to more fully explore the range of potential properties of a mechanical-thermal multifunctional metamaterial, compared with the current approach of using parametric studies. A methodology for constructing a design envelope using optimization is also presented. This approach can be applied to any type of metamaterial design, given a method for automatically generating and analysing any potential design (e.g. using scripting and numerical analysis). The approach is also general and can include other material properties, such as stiffness and density, and other constraints such as manufacturing limitations.

## 5. Conclusions

This paper demonstrates how optimization can be used to design a multifunctional metamaterial and more fully explore the design envelope of potential material properties. The example presented is a 2D mechanical-thermal metamaterial, based on a star-shaped lattice proposed by Ai and Gao (2017). This metamaterial can simultaneously exhibit both negative Poisson's ratio (auxetic behaviour) and non-positive thermal expansion coefficient (including near-zero values).

An optimization framework is developed that combines FEA-based periodic homogenization, implemented in Abaqus using the EasyPBC plugin, with a particle swarm-based optimizer from the pyOpt package. The framework is used to solve several optimization problems that explore the full potential of the studied multifunctional metamaterial. The results show that this optimization-based approach obtains a wider range of potential properties, compared with the previous study that used parametric studies. This is due to the optimizer being able to change all parameters simultaneously to find the best design, whereas parametric studies usually change one variable at a time, whilst keeping all others fixed.
This optimization-based approach is general and could be applied to better understand the potential of other metamaterials, existing or yet to be invented.

## Appendix A

Table 4. Full results for the design envelope in Fig. 6.

|  | $H_1$ | $H_2$ | θ (°) | t | PR, $v$ | NCTE |
|---|---|---|---|---|---|---|
| PR MAX | 47.7 | 5 | 39.56 | 4.74 | 0.0003 | 0.241 |
| NCTE MAX | 100 | 5 | 5 | 5 | -0.0044 | 0.285 |
| PR MIN (NCTE larger than set value) | 100 | 5 | 14.62 | 1.37 | -0.058 | 0.25 |
|  | 100 | 5 | 15.00 | 0.82 | -0.119 | 0.2 |
|  | 98.69 | 5.01 | 16.58 | 0.5 | -0.22 | 0.1 |
|  | 100 | 7.05 | 16.84 | 0.5 | -0.302 | -0.01 |
|  | 100 | 7.41 | 16.96 | 0.5 | -0.311 | -0.025 |
|  | 100 | 8.06 | 17.19 | 0.5 | -0.326 | -0.05 |
|  | 100 | 9.77 | 17.92 | 0.5 | -0.351 | -0.1 |
|  | 100 | 13.49 | 21.20 | 0.5 | -0.383 | -0.2 |
| PR MIN | 100 | 13.34 | 23.85 | 0.5 | -0.386 | -0.246 |
| PR MIN (NCTE larger than set value) | 100 | 13.99 | 26.51 | 0.5 | -0.383 | -0.3 |
|  | 100 | 14.70 | 31.12 | 0.5 | -0.365 | -0.4 |
|  | 100 | 15.11 | 35.24 | 0.5 | -0.342 | -0.5 |
|  | 100 | 15.41 | 38.90 | 0.5 | -0.317 | -0.6 |
| NCTE MIN | 100 | 25.01 | 40 | 0.5 | -0.291 | -0.647 |
| NCTE MIN | 100 | 46.31 | 40 | 0.5 | -0.25 | -0.638 |
|  | 100 | 83.16 | 40 | 0.5 | -0.2 | -0.621 |
| (PR larger than set value) | 69.7 | 100 | 40 | 0.5 | -0.15 | -0.604 |
| | 37.8 | 100 | 40 | 0.5 | -0.1 | -0.587 |
| | 15.9 | 100 | 40 | 0.5 | -0.05 | -0.569 |
| PR MAX (negative NCTE) | 5 | 100 | 40 | 5 | -0.015 | -0.551 |

**References**

Ai, L., and X-L. Gao. 2017. "Metamaterials with negative Poisson's ratio and non-positive thermal expansion." Composite Structures 162: 70-84.

Ai, L., and X-L. Gao. 2018. Three-dimensional metamaterials with a negative Poisson's ratio and a non-positive coefficient of thermal expansion. International Journal of Mechanical Sciences 135: 101-113.

Engheta, N., and R. W. Ziolkowski, eds. (2006) Metamaterials: physics and engineering explorations. John Wiley & Sons.

Geem, Z.W., J.H. Kim, G.V. Loganathan. 2001. "A new heuristic optimization algorithm: Harmony search." Simulation 76: 60–68.

Geers, M., Kouznetsova, V. and Brekelmans, W. 2010. "Multi-scale computational homogenization: Trends and challenges." Journal of Computational and Applied Mathematics, 234(7): 2175-2182.

Grima, J. N., P-S. Farrugia, R. Gatt, and V. Zammit. 2007. "Connected triangles exhibiting negative Poisson's ratios and negative thermal expansion." Journal of the Physical Society of Japan 76 (2): 025001.

Ha, C. S., E. Hestekin, J. Li, M. E. Plesha, and R. S. Lakes. 2015. "Controllable thermal expansion of large magnitude in chiral negative Poisson's ratio lattices." Physica status solidi (b) 252 (7): 1431-1434.

Huang, C., and L. Chen. 2016. "Negative Poisson's ratio in modern functional materials." Advanced Materials 28 (37): 8079-8096.

Jansen, P., and R. Perez. 2011. "Constrained Structural Design Optimization via a Parallel Augmented Lagrangian Particle Swarm Optimization Approach." International Journal of Computer and Structures, 89 (13-14): 1352–1366.

Lakes, R. 1987. Foam structures with a negative Poisson's ratio. Science 235: 1038-1041.

Lakes, R. 1996. "Cellular solid structures with unbounded thermal expansion." Journal of Materials Science Letters 15 (6): 475-477.

Li, X., L. Gao, W. Zhou, Y. Wang, and Y. Lu. 2019. "Novel 2D metamaterials with negative Poisson's ratio and negative thermal expansion." Extreme Mechanics Letters 30: 100498.

Ng, C. K., K. K. Saxena, R. Das and E. S. Flores. 2017. "On the anisotropic and negative thermal expansion from dual-material re-entrant-type cellular metamaterials." Journal of materials science 52 (2): 899-912.

Omairey, S.L., P.D. Dunning and S. Sriramula. 2019 "Development of an ABAQUS Plugin Tool for Periodic RVE Homogenisation". Engineering with Computers 35 (2): 567-577.

Perez, R. E., P. W. Jansen, & J.R.R. Martins. 2012. "pyOpt: a Python-based object-oriented framework for nonlinear constrained optimization." Structural and Multidisciplinary Optimization, 45 (1): 101-118.

Raminhos, J. S., J. P. Borges, and A. Velhinho. (2019). Development of polymeric anepectic meshes: auxetic metamaterials with negative thermal expansion. Smart Materials and Structures 28 (4): 045010.

Ren, X., R. Das, P. Tran, T. D. Ngo, and Y. M. Xie. 2018. Auxetic metamaterials and structures: A review. Smart materials and structures 27 (2): 023001.

Sigmund, O., and S. Torquato. 1996. "Composites with extremal thermal expansion coefficients." Applied Physics Letters 69 (21): 3203-3205.

Sklan, S. R., and B. Li. 2018. "Thermal metamaterials: functions and prospects." National Science Review 5 (2): 138-141.

Wei, K., Y. Peng, Z. Qu, Y. Pei, and D. Fang. 2018. "A cellular metastructure incorporating coupled negative thermal expansion and negative Poisson's ratio." International Journal of Solids and Structures 150: 255-267.